\documentclass[a4paper]{article}
\usepackage{graphicx}
\usepackage{amsmath}
\usepackage{amssymb}

\def\R{\mathbb{R}}

\def\N{\mathbb{N}}

\usepackage{bm}

\title{\bf Immersed boundary methods for numerical simulation of confined fluid and plasma turbulence in complex geometries: a review}
\author{Kai Schneider$^{1}$ \\ ~ \\
{\small $^1$M2P2-CNRS, Aix-Marseille Universit\'e} \\ 
{\small 38, Rue Fr\'ed\'eric Joliot-Curie, 13451 Marseille Cedex 13, France. } 
}
\date{\today}
\begin{document}

\maketitle

\begin{abstract}
Immersed boundary methods for computing confined fluid and plasma flows in complex geometries are reviewed.
The mathematical principle of the volume penalization technique is described and simple examples for imposing Dirichlet and Neumann boundary conditions in one dimension are given. Applications for fluid and plasma turbulence in two and three space dimensions illustrate the applicability and the efficiency of the method in computing flows in complex geometries, for example in toroidal geometries with asymmetric poloidal cross-sections.
\end{abstract}

\section{Introduction}

Immersed boundary techniques, including penalization approaches, are nowadays commonly employed to solve boundary or initial boundary value problems in complex geometries. They consist of embedding the original, possibly complex spatial domain inside a bigger domain having a simpler geometry, for example a Cartesian geometry, while keeping the boundary conditions approximately enforced thanks to new terms that are added to the equations.
Historically, the penalization technique can be traced back to Courant \cite{Cour43}  in the context
of constrained optimization. Later, Saulyev \cite{Saul63} applied it in the context of fictitious domain methods for immersed boundaries. 
In a classical paper~\cite{Pesk77}  an immersed boundary method was proposed for computing the blood flow in a beating heart.
Some history on immersed boundary techniques can be found in \cite{AKKS14}, and for detailed reviews we refer to the classical review papers \cite{Pesk02} and \cite{MiIa05}.

In the current work we focus on one particular example,  the volume penalization method \cite{ABF99} which, inspired by the physical intuition that a solid wall is similar to a vanishingly porous medium, uses the Brinkman-Darcy drag force as penalization term. The main advantage of such penalized equations is that they can be discretized independently of the geometry of the original problem, since the latter has been encoded into the penalization terms. Such a simplification permits a massive reduction in solver development time, since it avoids the issues associated with the design and management of the grid, allowing for example the use of simple fast Fourier transform (FFT)-based spectral solvers in Cartesian geometries. 
Using spectral solvers no linear systems have to be solved, e.g., to impose the incompressibility of the magnetic and velocity fields.
Furthermore the penalized equations are solved without introducing additional discretization errors, and thus effects of numerical diffusion and dispersion present in low-order numerical schemes are avoided. 
Moreover, parallel FFT libraries (e.g., P3DFFT)  are available on the current supercomputers.
The advantage of the penalization method becomes even more substantial when the geometry is time-dependent, as in the case of moving obstacles, or when fluid-structure interaction is taken into account.

In the following we give a non exhaustive overview of the development for different applications using volume penalization.
Two-dimensional hydrodynamic flows in staggered or inline tube bundles have been computed in \cite{Schne05}
and a dipole-wall collision benchmark using different numerical discretizations of the penalized Navier--Stokes equations has been obtained in \cite{KOKCSH07}.
Final states in two-dimensional decaying hydrodynamic turbulence in different geometries have been studied in \cite{ScFa08} and flows past flat plates focusing on the influence of the plate's geometry in \cite{SPVF14}.
Nguyen van yen et al. (2011) \cite{NKS14} computed dipole-wall collisions and derived the scaling of the energy dissipation in the large Reynolds number limit. They showed that it converges indeed to a finite value.
The penalized wave equation has been analysed in \cite{PCLS05}.
Applications to moving obstacles and fluid-structure interaction problems can be found in
\cite{KoSc09, KMFS11, KoES13} and \cite{EKSS13}.
In \cite{KSHPG13} a numerical study using the volume penalization method for impeller-driven von K\'arm\'an flows  has been performed.
Different simulations of compressible flows in complex geometries using the volume penalization
have been described in \cite{BCD09, LiVa07}.
An adaptive penalty method for incompressible Navier-Stokes has been proposed by Shirokoff and Nave \cite{ShNa13}.

%
The extension of penalization techniques to magnetohydrodynamic (MHD) flows is more recent.
Simulations of the self-organization of confined plasma in two space dimensions
studying the influence of the confinement geometry have been presented in \cite{BNS08,NBS08}.
The numerical method has been extended to three space dimensions and is described and benchmarked in detail in \cite{MLBS14}.
The spontaneous spin-up of plasma in toroidal geometries using the viscous-resistive MHD equations
has been discovered in Morales et al.~\cite{MBSM12}.
The effect of toroidicity in reversed field pinch devices is studied in \cite{MBSM14}.
Roberts et al. \cite{RLMBS14} investigated helically forced MHD flows in cylindrical geometries and showed the persistence of helical modes even when the dynamics becomes turbulent
A  penalty method for hyperbolic partial differential equations modeling the edge plasma transport in a tokamak
has been proposed in \cite{AAG14}.

The aim of this paper is to review the volume penalization method and to illustrate its potential for applications in fluid and plasma turbulence in complex domains.
Typically Dirichlet boundary conditions are considered corresponding to imposing the value of the solution at the boundary. Extensions for dealing with Neumann boundary conditions, i.e., imposing values of the derivative of the solution at the boundary, have been proposed in \cite{KKAS12,KNS14} based on previous work in \cite{RaB07}. 
An alternative approach for imposing either Dirichlet or Neumann boundary conditions which is based on sharp interface methods has been introduced in \cite{MDBN08}.
Considering simple examples in one space dimension allows an understanding and analysis of the convergence behavior of the penalization techniques when the penalization parameter tends to zero. We show that, for a given numerical discretization of the penalized equations, there exists a value of the penalization parameter, corresponding to a balance between penalization and discretization errors, below which no further gain in precision is
achieved. These results shed light on the behavior of volume penalization schemes when solving the penalized equations, outline the limitations of the method, and give indications on how to choose the penalization parameter at least in simple test cases. Nevertheless for practical applications a series of computations may be necessary to determine the actual value of the parameters, as discussed, e.g., in \cite{EKSS15}. Finally, different illustrations will be given for hydro- and magnetohydrodynamic problems in the turbulent regime including a study of the spatio-temporal self-organization of visco-resistive magnetohydrodynamics in toroidal geometries with different poloidal cross-sections while imposing curl-free toroidal magnetic and electric fields.

The outline of the paper is the following. First, we present a short primer on penalization.
Then we describe in some detail the volume penalization to impose either Dirichlet
or Neumann boundary conditions and we present some simple one-dimensional examples.
Applications to fluid and plasma turbulence in two and three dimensions illustrate the properties of the method.
Finally, we set out the conclusions together with some perspectives for future work.

\section{A short primer on penalization}

To illustrate the idea of volume penalization
we consider a boundary (BVP) or initial boundary value problem (IBVP) for a partial differential equation
in a domain $\Omega_f \subset \R^d$,
written in the following abstract operator equation
\begin{equation} \label{eqn:pde}
L u = f \quad {\rm for} \quad {\bf x} \in \Omega_f  
\end{equation}
completed with boundary conditions
$B u = g$ at $\partial \Omega_f$ and additional initial conditions in case of an IBVP.
The differential operator $L$ stands, for example, for the Laplace operator, $L = - \nabla^2$, or the Navier--Stokes
or Maxwell operator.
The boundary conditions can be of Dirichlet or Neumann type, e.g., $u = g$ or $\partial u / \partial n = g$, respectively, where $n$ is the outer normal of the domain.
Solving eq.~(\ref{eqn:pde}) numerically requires domain-fitted grids using, e.g., finite elements.
The grid generation and the numerical solution of the resulting discretized problem can be demanding,
see for example \cite{FePe96}.

An alternative are penalization methods which embed the problem posed in the complexly shaped domain $\Omega_f$  
into a larger simple domain, typically  of rectangular shape $\Omega$, i.e., $\Omega_f \subset \Omega$.
The advantage is that fast solvers are available for such problems, which can furthermore be easily parallelized.

In the Dirichlet case the penalized problem thus reads
\begin{equation} \label{eqn:pde_penal}
L u_\eta = f -  \frac{1}{\eta} \chi (B u - g)\quad {\rm for} \quad {\bf x} \in \Omega = \Omega_f \cup \Omega_s \, ,
\end{equation}
where the boundary conditions have been included into the equation and an additional parameter, the small penalization parameter $\eta> 0$ has been introduced.
All information about the geometry of $\Omega_f$ has been encoded into the mask function $\chi$, which is defined
as
\begin{equation}
\chi(\bf x) \, = \,   \left \{
    \begin{array} {ll}
        0 \quad \quad \quad \; \mbox{\rm for} \quad  {\bf x} \in \Omega_f\\
        1 \quad \quad \quad \; \mbox{elsewhere}\\
    \end{array}
  \right.
\label{eqn:mask}
\end{equation}
In the domain $\Omega_f$ the mask function vanishes identically and thus the original PDE in eq.~(\ref{eqn:pde}) is satisfied.

The difference between between the solution $u$ of the original PDE~(\ref{eqn:pde}) and the solution of the penalized problem $u_\eta$ (eq.~(\ref{eqn:pde_penal})) is the penalization (or modeling) error, which depends
on the size of the penalization parameter $\eta$,
\begin{equation} \label{eqn:error_penal}
 \parallel u - u_\eta \parallel \propto \eta^\alpha
\end{equation}
where $\alpha$ describes the order of the penalization method and $\parallel \cdot \parallel$ is a suitable norm,
e.g., the energy norm $\parallel f \parallel_2 = ( \int_{\Omega_f} |f({\bm x})|^2 d{\bm x} )^{1/2}$ or the $L^1$ or $L^\infty$ norm.
For convergence of the method the solution of the penalized problem $u_\eta$ should tend towards
the solution of the original problem $u$, i.e., ${\rm lim}_{\eta \rightarrow 0} \parallel u - u_\eta \parallel \rightarrow 0$.
Thus $\alpha > 0$ is required for convergence of the method.
For the volume penalization we have $\alpha= 1/2$ \cite{ABF99,CaFa03}.

Applying a numerical method to the penalized PDE eq.~(\ref{eqn:pde_penal})
we obtain the discretized penalized equation,
\begin{equation} \label{eqn:pde_penal_discret}
L^N u^N_\eta = f^N -  \frac{1}{\eta} \chi^N (B u - g) \quad {\rm for} \quad {\bf x} \in \Omega
\end{equation}
with $\bigtriangleup x \propto 1/N$, where $N$ denotes the number of grid points and $L^N$ is the discretized version
of the operator $L$.
In the case of an IBVP a suitable time discretization is also necessary.
Explicit time discretization of the penalization term implies a time step restriction for stability.
Typically the time step is limited by $\eta$ and hence the problem becomes stiff in the limit $\eta \rightarrow 0$.

The solution of the discretized penalized equations $u^N_\eta$, given that the discretization is consistent and the numerical scheme is stable, converges towards the exact solution  $u_\eta$ of the penalized equation (\ref{eqn:pde_penal}).
The discretization error,
\begin{equation} \label{eqn:disc_error}
 \parallel u_\eta - u^N_\eta \parallel \propto \left( \frac{1}{N} \right)^\beta
\end{equation}
depends not only on the order of the numerical scheme, but is also limited by the regularity, i.e., the smoothness of the exact solution $u_\eta$ of the penalized problem.
The order $\beta$ is thus determined as the mimum of the order of the underlying numerical scheme and the
regularity of the exact penalized solution.

\medskip

Finally, the question in  computations using the penalization method concerns the error between the exact solution $u$
of the BVP (or IBVP) given by eq.~(\ref{eqn:pde}) and the numerical solution of the penalized equation eq.~(\ref{eqn:pde_penal_discret}).
This total error has two contributions: the modeling error which is due to the penalization; and the discretization error
which is due to the numerical solution of the discretized problem.
Applying the triangle inequality the following estimate holds:
\begin{equation} \label{eqn:tot_error}
\parallel u - u^N_\eta \parallel \; \leq \;   \parallel u - u_\eta \parallel + \parallel u_\eta - u^N_\eta \parallel
\end{equation}
and thus we obtain a bound for the total error.
A straightforward argument would suggest choosing very small values for $\eta$ to minimize the modeling error.
However this would imply that the problem becomes stiff, one loses regularity in the exact solution of the
penalized problem and the leading-order constant of the discretization error blows up and a very fine grid would become necessary.
Thus to optimize the estimate both errors should be of the same order of magnitude.
This shows that the penalization parameter $\eta$ and the numerical resolution $N$ of the discretized problem are coupled and should be chosen accordingly. For further discussion we refer to \cite{NKS14}.

\begin{figure}[htbp]
\begin{center}
\includegraphics[scale=0.2]{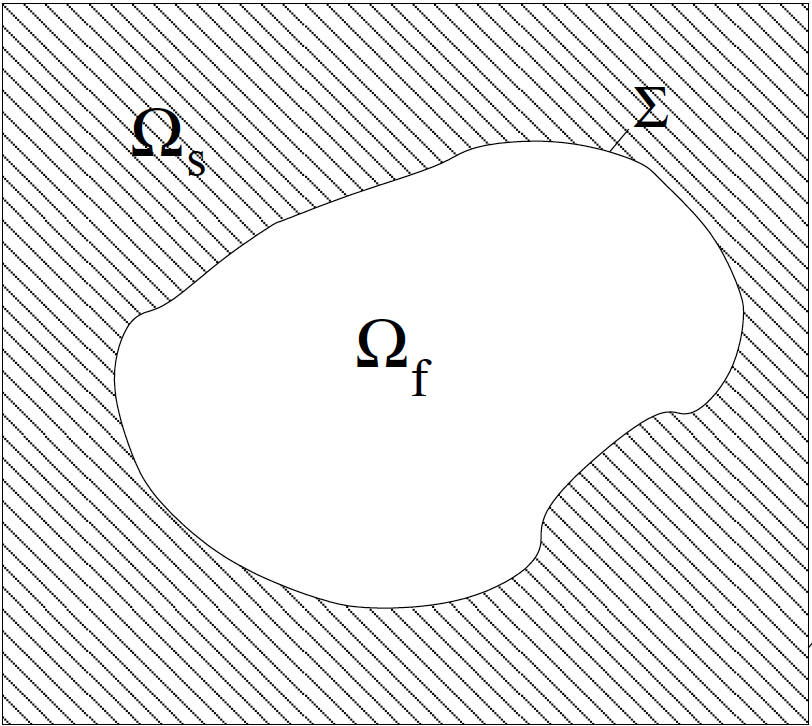}
\end{center}
\caption{Sketch of the fluid domain $\Omega_f$ immersed into the solid domain $\Omega_s$. The boundary of the fluid domain is denoted by $\Sigma = \partial \Omega_f$. The total computational domain is $\Omega = \Omega_f \cup \Omega_s$.}
\label{fig:sketch}
\end{figure}

\section{Volume penalization for Dirichlet and Neumann boundary conditions}

For illustration we consider first a simple toy problem, the Poisson equation in one space dimension,
\begin{equation} \label{eqn:poisson1d}
- \frac{d^2u}{dx^2} \; = \; f \quad {\rm for} \quad { x} \in \Omega_f 
\end{equation}
completed with either homogeneous Dirichlet ($u = 0$ for $x \in \partial \Omega_f$) or Neumann boundary conditions ($du/dx = 0$ for $x \in \partial \Omega_f$).
We choose $\Omega_f = ] 0,\pi [ $ and for the right-hand side $f$ a trigonometric function, i.e., $f(x) = m^2 \sin mx$
and $f(x) = m^2 \cos mx$, respectively, with $m \in \N$.
The exact solution is $u(x) = \sin mx$ and $u(x) = \cos mx +C$  in the Dirichlet and Neumann case, respectively.
The additive constant $C \in \R$ shows that the solution is not unique for Neumann boundary conditions.
To guarantee in this case the existence of a solution, the right-hand side $f$ has to satisfy the compatibility condition,
$\int_{0}^{\pi} f(x) dx = u'(x=\pi) - u'(x=0) = 0$.

\subsection{The penalized Poisson equation}

Now we replace the original problems by the penalized problems.
The domain $\Omega_f = ] 0, \pi [ $ is embedded in the larger domain $\Omega = ] 0, 2 \pi [$.
Thus we have $\Omega = \Omega_f \cup \Omega_s$, where $\Omega_s$ is the penalization domain.
For further simplification we impose periodic boundary conditions at the boundary of $\Omega$.

In the Dirichlet case we obtain the penalized Poisson equation
\begin{equation} \label{eqn:poisson1d_penalized_dirichlet}
- \frac{d^2 u_\eta}{dx^2} \; + \frac{1}{\eta} \;  \chi \, u_\eta \; = \; f  \quad \mbox{\rm for} \quad x \in ] 0, 2 \pi [
\end{equation}
while in the Neumann case we have
\begin{equation}
- \frac{d}{dx} \left( (1 - \chi) + \eta \chi \right) \frac{d}{dx} u_\eta  \, = \, f \quad \mbox{\rm for} \quad x \in ] 0, 2 \pi [
\label{eqn:poisson1d_penalized_neumann}
\end{equation}
The penalization term in the Dirichlet case forces the solution to vanish inside the penalization domain and thus imposes homogeneous Dirichlet conditions at the interface.
In the Neumann case the boundary condition corresponds to zero flux through the interface of $\Omega_f$ and $\Omega_s $. This can be achieved by imposing vanishing diffusivity (of order $\eta$) inside the penalization domain.
The function $f$ is extended in the larger domain by zero padding, i.e., $f(x) = 0$ for $x\ \in \Omega_s$.
Again both problems can be solved analytically in each sub-domain and we obtain
in the Dirichlet case \cite{NKS14}
\begin{equation}
u_\eta(x) \, = \,   \left \{
    \begin{array} {ll}
        \sin m x + A_1 x + A_2 \quad \mbox{\rm for} \quad x \in ]0, \pi [ \\
        \frac{m^2 \eta}{1 + \eta m^2} \sin mx + B_1 \exp(-x/\sqrt \eta) + B_2 \exp(x/\sqrt \eta)    \quad  \mbox{\rm for} \quad x \in ]\pi, 2 \pi [
    \end{array}
  \right.
\end{equation}
and in the Neumann case \cite{KNS14}
\begin{equation}
u_\eta(x) \, = \,   \left \{
    \begin{array} {ll}
        \cos m x + A_1 x + A_2 \quad \mbox{\rm for} \quad x \in ]0, \pi [ \\
        B_1 x + B_2            \quad \quad \quad \quad \quad \mbox{\rm for} \quad x \in ]\pi, 2 \pi [
    \end{array}
  \right.
\end{equation}
Imposing continuity of the solution and of the derivative at $x=0 = 2\pi$ and $x=\pi$ the coefficients ($A_1, A_2, B_1, B_2$) can be determined.
The corresponding values can be found in \cite{NKS14, KNS14}.
Note that in the Neumann case only three out of the four coefficients can be determined as the solution is not unique.
The penalization error $\parallel u - u_\eta \parallel$ can thus be explicitly computed and we find in the
Dirichlet case the expected $O(\sqrt \eta)$ behavior, while in the Neumann case an $O(\eta)$ behavior is found
which is better than $O(\sqrt \eta)$ shown in \cite{KKAS12} for the heat equation.
In Fig.~\ref{fig_error_1dpoisson} (top) the exact solution of the penalized Dirichlet problem is plotted for two values of $\eta$ (left) together with the first derivative (right). We find that for decreasing $\eta$ the penalized solution tends towards the solution of the unpenalized problem. We also observe the existence of a boundary layer in the penalized domain (at the interface of $\Omega_f$ and $\Omega_s$, close to $x=\pi$) which becomes steeper when $\eta$ becomes smaller.
For $\eta = 10^{-6}$ we find that the first derivative becomes almost discontinuous at $x= \pi$. This shows that the regularity of $u_\eta$ is lost in the limit of small $\eta$ and the problem becomes stiff.

\subsection{The discretized penalized Poisson equation}

To solve the penalized equations numerically we apply a second-order finite difference discretization to the penalized problems (\ref{eqn:poisson1d_penalized_dirichlet}) and (\ref{eqn:poisson1d_penalized_neumann}).
The computational domain $\Omega = [0, 2 \pi]$ is discretized with $N$ grid points $x_i = i/(2 \pi), i = 0, ..., N-1$
applying periodic boundary conditions.
We obtain the following linear system in the Dirichlet case
\begin{equation}
\left(-D_2 +\frac{1}{\eta} \vec \chi I \right) U = F 
\label{poisson1d_penalized_linsystem_dirichlet}
\end{equation}
with the vectors $U = (u(x_0), \dots, u(x_{N-1}))$, $\vec \chi = (\chi(x_0), \dots, \chi(x_{N-1}))$, $F = (f(x_0), \dots, f(x_{N-1}))$  in $\R^N$,  the identity matrix $I$ and where
\begin{equation}
  D_{2} = \frac{1}{h^2} \left(\begin{array}{ccccc}
    -2  &  1  &    &  & 1 \\
    1  & -2 &1  &  & \\
    ~  &     &   & \ddots & \\
        &     & 1  & -2   &  1  \\
      1 &    &   & 1 & -2 \\
  \end{array}\right),
\label{eq:d2_fd2}
\end{equation}
is the second-order central finite difference operator with $h = 2 \pi /N$.
The resulting tridiagonal system can be solved using standard numerical linear algebra tools.

In the Neumann case we get a second-order approximation with the following finite difference scheme,
\begin{equation}
 -\frac{1}{2} \left( D_F \Theta D_B + D_B \Theta D_F \right)   \, U \; = \; F
\label{poisson1d_penalized_linsystem_neumann}
\end{equation}
where
$\Theta = [\theta(x_0), \theta(x_1), ...,  \theta(x_{N-1})]$
with $\theta(x_i) = 1 - \chi(x_i) + \eta \chi(x_i))$.
The first order backward and forward finite difference operators $D_B$ and $D_F$ are defined respectively, by
\begin{equation}
  D_{B} = \frac{1}{h} \left(\begin{array}{ccccc}
    1  &   &   & & -1 \\
    -1 & 1 &   & & \\
    ~  &   &   & \ddots & \\
       &   &   & -1 & 1 \\
  \end{array}\right),
\quad\quad
  D_{F} = \frac{1}{h} \left(\begin{array}{ccccc}
    -1 & 1  &   & & \\
       & -1 & 1 & & \\
    ~  &    &   & \ddots & \\
    1  &    &   & & -1 \\
  \end{array}\right)
\label{eq:df_db_fd2}
\end{equation}
The linear system (\ref{poisson1d_penalized_linsystem_neumann}) is singular, the matrix has an eigenvalue $0$ and a solution only exists if the right-hand side $F$ is in its image.
Special care has to be taken for solving the linear system using either the pseudo-inverse, or removing one equation.

\begin{figure}[htbp]
\begin{center}
\includegraphics[width=\linewidth]{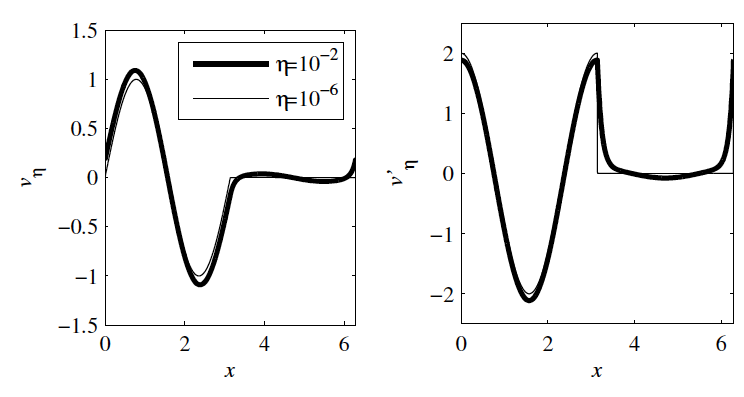} \\
\includegraphics[width=0.5\linewidth]{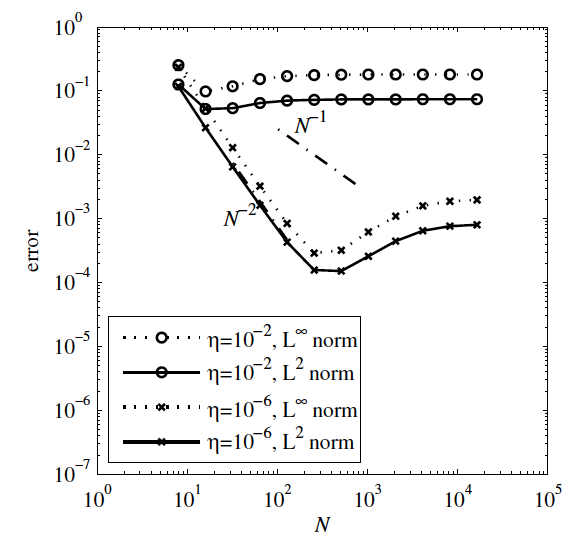}
\end{center}
\caption{\label{fig_error_1dpoisson} Exact solution $u_\eta$ (top, left) and its first derivative $u'_\eta$ (top, right) of the penalized Poisson equation with Dirichlet boundary condtions for $\eta =10^{-2}$ and $10^{-6}$. 
Convergence of the second-order finite difference scheme. Error with respect to the exact solution of the Dirichlet problem in $\Omega_f$ (bottom). From~\cite{NKS14}.}
\end{figure}

Performing numerical experiments \cite{NKS14, KNS14} showed that  second-order convergence of the numerical solution of the penalized equations towards the exact solution can be obtained, under the condition that $\eta$ is sufficiently small.
Moreover, for fixed $\eta$ the total error (modeling plus discretization error) has a minimum at $N \propto 1/\sqrt \eta$.
Fig.~\ref{fig_error_1dpoisson} (bottom) illustrates this for the Dirichlet case.
Varying $\eta$ with $1/N$ like $\eta \propto 1/N^2$ thus yields the best results in terms of minimum total error~\cite{NKS14}.
Nevertheless it must be mentioned that for spectral or fourth-order finite difference discretizations only first-order convergence can be observed. The explanation is that the linear system using second-order finite differences
is exactly the same for the penalized problem as for the original Dirichlet problem at all points inside the 
fluid domain, except at $x_1$ and $x_{N/2 -1}$ where values of $O(\sqrt \eta)$ are involved~\cite{NKS14}.
Hence for sufficiently small values of $\eta$ second-order convergence can be found.
Note that for fourth order finite differences the stencil is larger, thus more grid points are affected by the penalization term and the second-order convergence is lost.

\subsection{The penalized Navier--Stokes and Maxwell equations}

Analogously to the penalized Poisson equation we can impose no slip and no penetration, i.e., Dirichlet boundary conditions in the incompressible Navier--Stokes equations by adding a penalization term to the momentum equation,
\begin{eqnarray}
\frac{\partial {\bf u}}{\partial t} \, + \, {\bm \omega} \times {\bf u}  \, + \, \nabla \Pi  \, - \, \nu \nabla^2 {\bf u} \, &=& \, - \frac{1}{\eta} \chi ({\bf u} - {\bf u}_{wall}) \\
\nabla \cdot {\bf u} \; & =& \; 0 \nonumber
\label{eqn_ns_penalized}
\end{eqnarray}
where ${\bm \omega} = \nabla \times {\bf u}$ is the vorticity and $\Pi = p + {\bf u}^2/2$ is the modified pressure
which satisfies $- \nabla^2 \Pi = \nabla \cdot ( {\bm \omega} \times {\bf u}  + \frac{1}{\eta} \chi ({\bf u} - {\bf u}_{wall}))$.
The kinematic viscosity is denoted by $\nu$.
The velocity at the wall, ${\bf u}_{wall}$ does vanish identically in the case of fixed walls, or is prescribed in the case of moving walls.
In the limit $\eta \rightarrow 0$ the solution of the penalized Navier--Stokes equations tends to the solution
of the Navier--Stokes equations with no-slip boundary conditions and the penalization error is of order $O(\sqrt{\eta})$,
as shown in \cite{ABF99,CaFa03}.

The induction equation which describes the evolution of the magnetic field $\bf B$ can be penalized in a similar way
and we obtain
\begin{eqnarray}
\frac{\partial {\bf B}}{\partial t} \, - \, \nabla \times ( {\bf u} \times {\bf B})  \,   - \, \lambda \nabla^2 {\bf B} \, &=& \, - \frac{1}{\eta} \chi ({\bf B} - {\bf B}_{wall}) \\
\nabla \cdot {\bf B} \; & =& \; 0 \nonumber
\label{eqn_induction_penalized}
\end{eqnarray}
where $\lambda$ is the magnetic diffusivity.
The magnetic field at the wall, $ {\bf B}_{wall}$,  can be freely chosen and hence not all components have to be penalized. 
For example choosing ${\bf B}_{wall} = B_{\parallel}$ where $B_\parallel$ is the component of $\bf B$ parallel to the wall, only penalizes the normal component and leaves the parallel component free.
This allows modeling perfectly conducting boundary conditions.
Note that in the MHD case no mathematically rigorous convergence theorem has been proven yet, but in \cite{MLBS14}
asymptotic arguments for estimating the penalization error have been given. 

For the transport of a passive scalar $\xi$ we consider the advection-diffusion equation where no-flux conditions $\nabla \xi \cdot n =0$, i.e., homogeneous Neumann boundary conditions are imposed \cite{KKAS12},
\begin{equation}
\frac{\partial \xi}{\partial t} \, + \, [(1 - \chi) {\bf u}] \cdot \nabla \xi \, = \, \nabla \cdot [\kappa (1 - \chi) + \eta_\xi \chi] \nabla \xi
\label{eqn_advec_diff_penalized}
\end{equation}
where $\eta_\xi$ is the penalization parameter for the scalar $\xi$ and $\kappa$ its diffusivity.
In the limit $\eta_\xi \rightarrow 0$ the solution of the penalized equation tends to the solution of the advection-diffusion equation with no-flux boundary conditions and the penalization error is of order $O(\sqrt{\eta})$,
as shown in \cite{KKAS12}.

\section{Application to fluid turbulence}

In the following we consider different applications to two- and three dimensional hydrodynamic incompressible flows with no-slip walls and also the transport of a passive scalar where no-flux boundary conditions are imposed.
The numerical method is based on a pseudo-spectral discretization of the penalized Navier--Stokes and advection-diffusion equations and for details we refer the reader to~\cite{Schne05,KoSc09,KKAS12,EKSS15}. 

\subsection{Two-dimensional turbulence in a circular container}

\begin{figure}[htbp]
\begin{center}
\includegraphics[width=\linewidth]{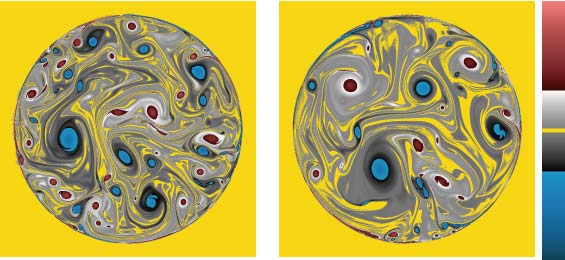}
\end{center}
\caption{\label{fig_2dvortcirc}Decaying two-dimensional flow in a container with no-slip boundary conditions at initial Reynolds number $5  \times 10^4$. Snapshots of vorticity $\omega$. From \cite{ScFa05}.}
\end{figure}

\begin{figure}[htbp]
\begin{center}

\includegraphics[height= 7cm,width=0.7\linewidth]{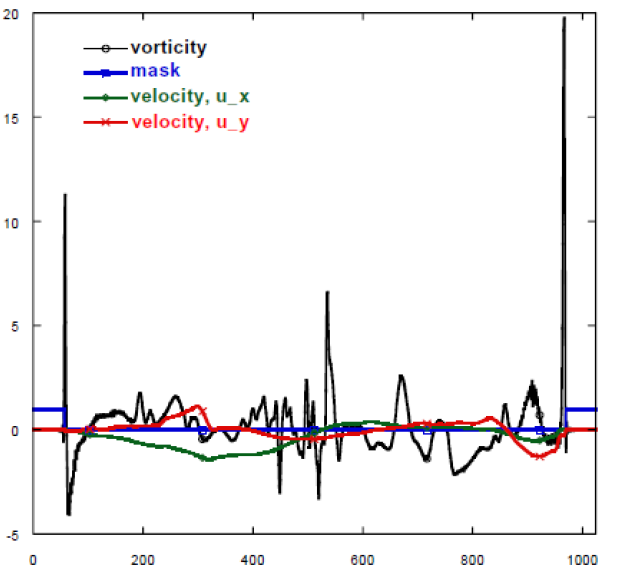} \\ \vspace{1cm}
\includegraphics[height=7cm,width=0.7\linewidth]{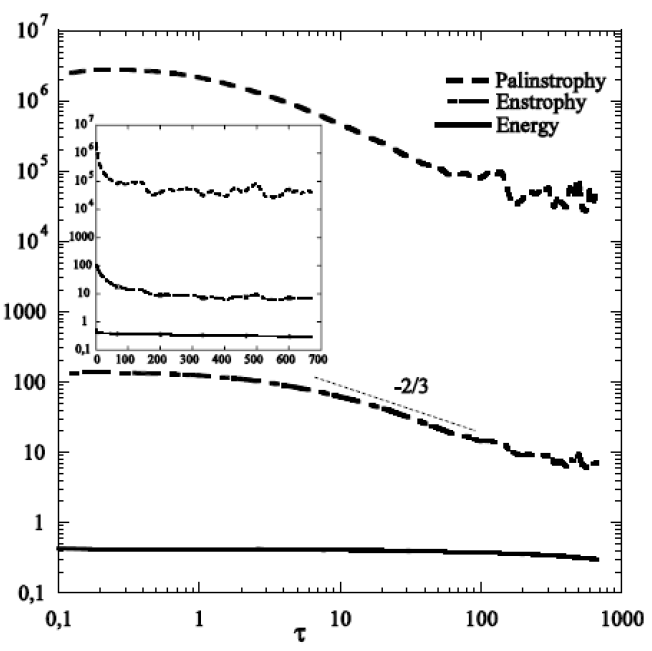}
\end{center}
\caption{\label{fig_2dvortcirc_int} Decaying two-dimensional container with no-slip boundary conditions at initial Reynolds number $5  \times 10^4$. 1d vertical cuts of vorticity, the two velocity components and the mask function (top). 
Evolution of the kinetic energy $E(t) = \frac{1}{2} \int |{\bm u}|^2 d{\bm x}$, enstrophy $Z(t) = \frac{1}{2} \int |{\omega}|^2 d{\bm x}$ and palinstrophy $P(t) = \frac{1}{2} \int |{\nabla \omega}|^2 d{\bm x}$ in double logarithmic representation (bottom), where $\tau = t / t_e$ is based on the initial eddy turn over time $t_e = \sqrt{2 Z(t+0)} = 0.061$. The flow has been integrated for $650 t_e$, corresponding to more than $10^5$ time steps. The inset shows the corresponding log-lin representation.  From \cite{ScFa05}.}
\end{figure}

In \cite{ScFa05} we presented numerical simulations of two-dimensional decaying turbulence 
in a circular container with no-slip boundary conditions computed at resolution $1024^2$ using the volume penalization technique.
Starting with random initial conditions with Reynolds number $Re = 5  \times 10^4$ where $Re$ is based on the domain size, the turbulent kinetic energy and the kinematic viscosity $\nu$, the flow rapidly exhibits self-organization into coherent vortices. Two snapshots of the vorticity field  $\omega$ at later times are shown in Fig.~\ref{fig_2dvortcirc}.
One-dimensional cuts in Fig.~\ref{fig_2dvortcirc_int} (top) illustrate the intermittent character of the vorticity field and the spikes at the domain boundary show the strong production of vorticity in a thin boundary layer due to the no-slip boundary conditions.
The cut of the mask function together with the cuts of the velocity components confirm that in the penalization domain the velocity does indeed vanish.
The formation of coherent vortices and the viscous boundary layer have significant impact on the production and decay of integral quantities. 
The evolution of kinetic energy $E(t)$, enstrophy $Z(t)$ and palinstrophy $P(t)$ are shown in Fig.~\ref{fig_2dvortcirc_int} (bottom).
The corresponding balance equations read,
\begin{equation}
d_t E = - 2 \nu Z \, , \quad d_t Z = - 2 \nu P \, + \, \oint_{\partial \Omega} \, \omega \, ({\bm n} \cdot \nabla \omega) \, ds \, , 
\end{equation}
where $\bm n$ denotes the outer normal with respect to $\partial \Omega$.
The source term in the enstrophy dissipation equation involves the vorticity and its gradient at the boundary
and yields a significant contribution in the small viscosity limit.
The no-slip wall produces vortices which are injected into the bulk flow and tend to compensate the enstrophy dissipation as observed in Fig.~\ref{fig_2dvortcirc_int}.
%
%
%
The self-organization of the flow is also reflected by the transition of the initially Gaussian vorticity probability
density function (PDF) towards a distribution with exponential tails. Because of the presence of coherent
vortices the pressure PDF becomes strongly skewed with exponential tails for negative values.
Details can be found in~\cite{ScFa05}.
%

\subsection{Passive scalar transport in two-dimensional confined turbulence}

To illustrate the volume penalization for imposing no-flux boundary conditions
we show a numerical simulation of a flow with a passive scalar in a simplified mixing device \cite{KKAS12}.
The penalized Navier--Stokes equations (\ref{eqn_ns_penalized}) are solved together with the penalized advection-diffusion equation (\ref{eqn_advec_diff_penalized}) in a square domain with periodic boundary conditions using a Fourier pseudo-spectral method.
The fluid domain corresponds to a circular vessel in which a cross-shaped rotor in the center of the domain rotates in the clockwise direction.
The boundary conditions are no slip for the velocity and no flux for the passive scalar.
%
\begin{figure}[htbp]
\begin{center}
\includegraphics[width=0.49\linewidth]{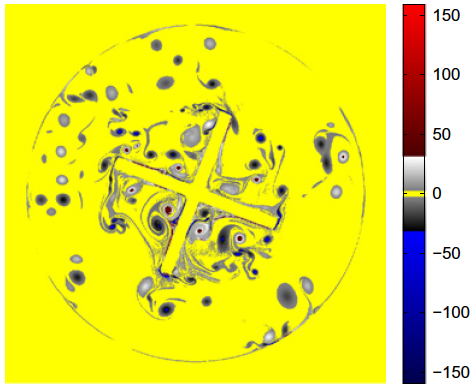}
\includegraphics[width=0.49\linewidth]{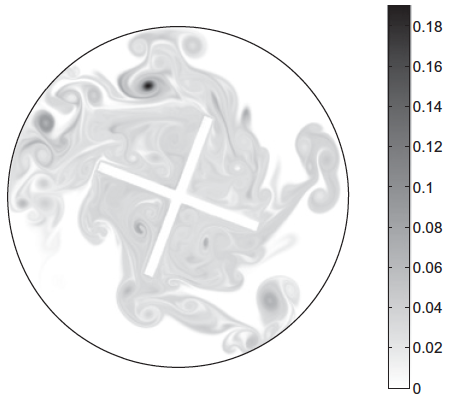}
\end{center}
\caption{\label{fig_2dpass_scal} Vorticity field in a circular vessel with a cross-shaped clockwise rotating rotor (left). Corresponding passive scalar field (right). From~\cite{KKAS12}.}
\end{figure}
%
The vorticity field  in Fig.~\ref{fig_2dpass_scal} illustrates the formation of boundary layers at the wall and at the rotor.
The formed vortex sheets destabilize and detach forming coherent vortices which are ejected into the bulk flow.
The corresponding passive scalar field, the initial condition corresponds to a Gaussian blob, is advected by the mean rotation induced by the rotor. The mixing process is further enhanced by the coherent vortices generated by the rotor blades and the boundary layer of the vessel.
For further details we refer the reader to \cite{KKAS12}.

\subsection{Flow past flexible flapping plates in  three dimensions}

Motivated by simplified models for swimming organisms or robots, which rely
on chord-wise flexible elastic plates, we present numerical simulations of fluid-structure interaction
using the volume penalization.
We consider a plate made out of linearly elastic inextensible material, which is perfectly rigid in the spanwise direction but flexible in the chordwise direction.
The plate can be thus modeled by the nonlinear beam equation with clamped free boundary conditions and it is solved with classical finite differences. The fluid part is solved with a pseudo-spectral method and volume penalization \cite{EKSS15}. 
Depending on the fluid/plate density ratio up to 25 iterations of the fluid-solid coupling are necessary within each time step. Details on the numerical method are described in \cite{EKSS15}.

\begin{figure}[htbp]
\begin{center}
\includegraphics[height=4cm,width=0.32\linewidth]{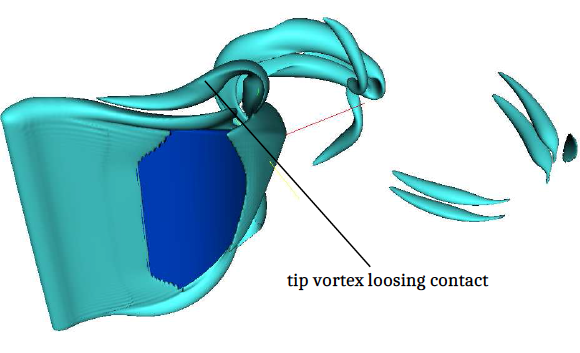}
\includegraphics[height=4cm, width=0.32\linewidth]{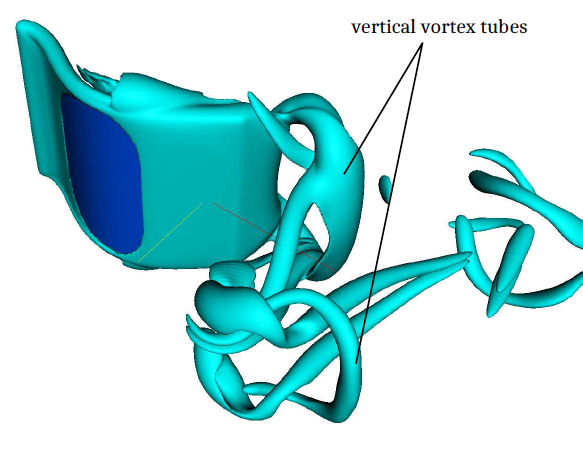}
\includegraphics[height=4cm,width=0.32\linewidth]{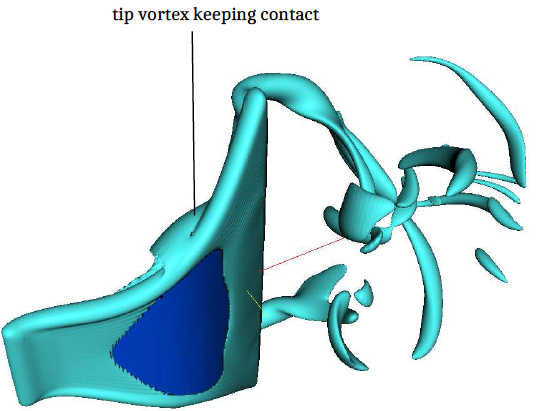}
\end{center}
\caption{\label{fig_plate3d} Flow generated by flapping cord-wise flexible plates of contracting (left), rectangular (center) and expanding shape (right) with imposed pitching motion at their leading edge and imposed mean flow (from left to right). Shown are isosurfaces of vorticity $\parallel {\bf \omega} \parallel = 17.5$. From~\cite{EKSS14}.}
\end{figure}

In \cite{EKSS14} we considered a configuration with imposed mean flow, and imposed at the leading edge of the plate a sinusoidal pitching motion. The Reynolds number is about $Re \approx 1000$.
We first simulated a swimmer with a rectangular plate and compared the results with a recent experimental study, before considering also an expanding and a contracting shape of the plate. Flow visualizations for the three geometries showing vorticity isosurfaces are given in Fig.~\ref{fig_plate3d}.
The tip vortices observed in Fig.~\ref{fig_plate3d} originate from three-dimensional effects due to the finite span. These 
have important effects for predicting the swimmer’s cruising velocity, since they contribute significantly to the drag force. 
We found that the cruising velocity of the contracting swimmer is larger than of the rectangular one, which in turn is larger than the expanding one. This observation can be explained by the relative importance of the tip vortices which interact differently with the flexible plates for the three considered geometries of the swimmer.
For the contracting case the tip vortices rapidly detach and thus reduce drag, while in the expanding case they are attached down to the trailing edge.

\section{Applications to plasma turbulence}

%
For the numerical simulation of plasma turbulence we consider the non-ideal MHD equations (eqs.~\ref{eqn_ns_penalized} and \ref{eqn_induction_penalized}) in which both viscous and resistive effects are taken into account. 
The magnetic Prandtl number, defined as the ratio between kinematic viscosity $\nu$ and magnetic diffusivity $\lambda$, is equal to one. 
An isothermal, incompressible plasma is considered with uniform and constant transport coefficients.
This approximation simplfies the problem as much as possible, while retaining the required level of complexity to study the nonlinear dynamics.
The boundary conditions corresponding to solid domains which are perfect conductors are imposed with the volume penalization method.
The numerical code is based on a Fourier pseudo-spectral discretization using FFTs to compute the derivatives and to solve the Poisson equations. It is described in detail in \cite{MLBS14}, including benchmarking and detailed validation studies.

\subsection{Spin-up in two-dimensional confined MHD}

We consider first two-dimensional decaying MHD turbulence in bounded domains and investigate
the spontaneous self-organization with a particular emphasis on the symmetry-breaking
induced by the shape of the confining boundaries; for details we refer to \cite{BNS08} and \cite{NBS08}. 
This symmetry-breaking is quantified by the angular momentum, which is shown to be generated rapidly and spontaneously from initial conditions free from angular momentum when the geometry lacks axisymmetry. 
In \cite{BNS10} this effect was illustrated by considering circular, square, and elliptical boundaries. It was shown that the generation of angular momentum in non-axisymmetric geometries can be enhanced by increasing the magnetic
pressure. Moreover, the effect becomes stronger at higher Reynolds numbers, which are based on the turbulent kinetic energy, the length scale of the domain and the kinematic viscosity. The generation of magnetic
angular momentum or angular field, previously observed in \cite{BNS08} at low Reynolds numbers, becomes
weaker at larger Reynolds numbers.
For axisymmetric geometries, the generation of angular
momentum is absent; nevertheless, a weak magnetic field can be observed. 
The derived evolution equations for both the angular momentum and angular field yield possible explanations for the observed behavior.

\begin{figure}[htbp!]
\begin{center}
\includegraphics[height=8cm,width=0.7\linewidth]{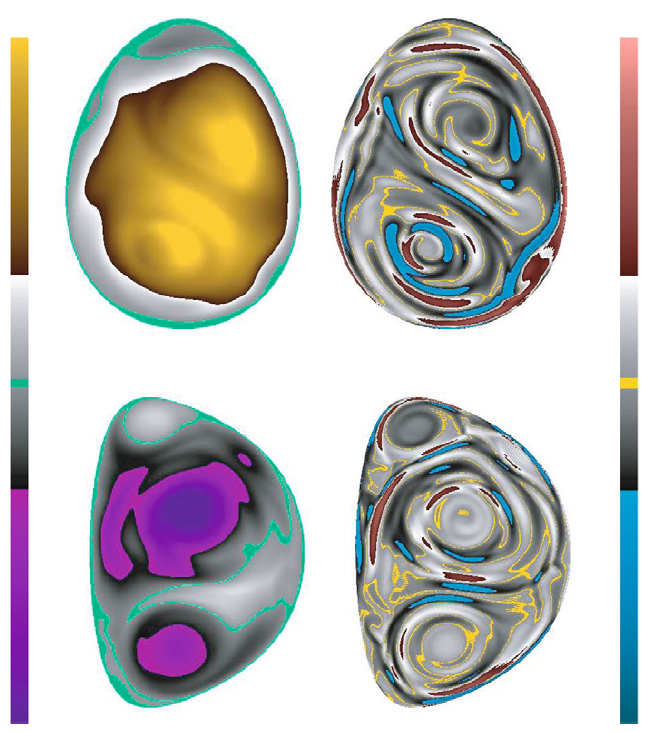}
\end{center}
\caption{\label{fig_spinup2d} Two-dimensional magnetized plasma: stream function (left) and vorticity
(right) for the ovoid (top) and the D-shaped geometry (bottom). The visualizations correspond to the time-instants at which the absolute
value of the angular momentum reaches its maximum. From~\cite{SNB11}.}
\end{figure}

\begin{figure}[htbp!]
\begin{center}
\includegraphics[width=0.6\linewidth]{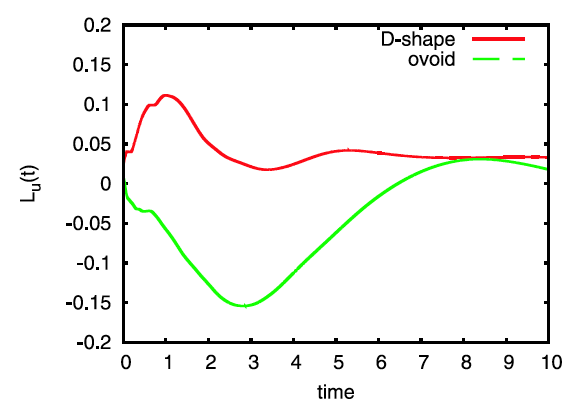}
\end{center}
\caption{\label{fig_spinup2d_momentum} Two-dimensional magnetized plasma: Time evolution of the angular momentum $L_u (t)$ for the ovoid and the D-shaped geometry. From~\cite{SNB11}.}
\end{figure}

In Fig. ~\ref{fig_spinup2d} we show simulations of two-dimensional decaying
MHD turbulence inside an ovoid and in a D-shaped geometry starting
with random initial conditions \cite{SNB11}. For both geometries we observe
that the two-dimensional magnetized plasma self-organizes into
a state containing large-scale flow structures, illustrated by the
stream function in Fig.~\ref{fig_spinup2d} (left) and vorticity, Fig.~\ref{fig_spinup2d} (right).
To quantify the spin-up we consider the angular momentum
defined as
\begin{equation}
L_u(t) \, = \, \int_{\Omega_f} {\bf e_z}  \cdot {\bf r} \times {\bf u} \, ds \, = \, 2 \int \psi \, ds
\end{equation}
$L_u$ quantifies the fluid rotation. 
The maximum angular momentum for a given kinetic energy is obtained for a fluid in solid body rotation. 
In this particular realization we find that the generation of angular
momentum is stronger in the ovoid than in the D-shaped geometry,
as shown in Fig.~\ref{fig_spinup2d_momentum}.

\subsection{Self-organization of confined MHD flows in toroidal domains}

\begin{figure}[h!]
\begin{center}
\includegraphics[width=\linewidth]{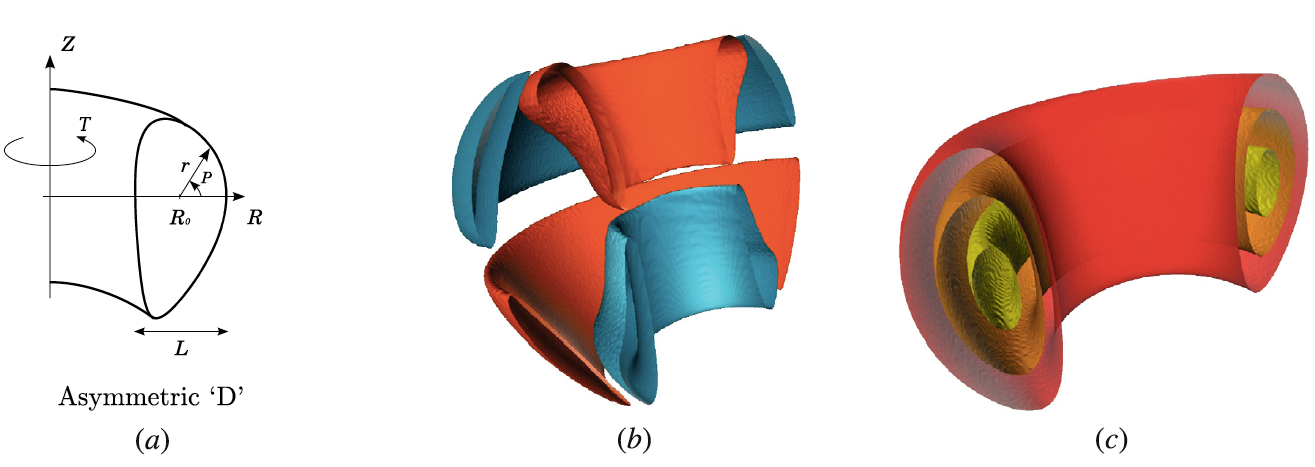}
\end{center}
\caption{\label{fig_torusasym_jcp}(a) Sketch of the toroidal geometry with asymmetric poloidal cross-section,
(b) toroidal velocity field component isocontours (blue $+9 \cdot 10^{−4}$, orange $−9 \cdot 10^{−4}$)
and (c) perturbed toroidal magnetic field component isocontours (red $+0 .025$, orange $+ 0.04$, yellow $+0.05$). From \cite{MLBS14}.}
\end{figure}

The spatio-temporal self-organization of visco-resistive magnetohydrodynamics in a toroidal geometry (see Fig.~\ref{fig_torusasym_jcp}, left) was studied in \cite{MBSM12} using fully three-dimensional simulations
considering two geometries: a torus with a symmetric poloidal cross-section and one with an asymmetric poloidal cross-section. 
The magnetized plasma is initially in a quiescent state and curl-free toroidal magnetic and electric fields are imposed.
The simulations show that spontaneously a flow field is generated in both geometries and the magnetized plasma starts to move. 
Moreover, the flow evolves from dominantly poloidal to toroidal when the Lundquist (or Reynolds) numbers $M$ are increased. 
Here the viscous Lundquist number $M$ is defined as $M = C_A L / \nu$ where $C_A$ is the toroidal Alfv\'en velocity,
$L$ the diameter of the cross section of the torus and $\nu$ the kinematic viscosity.
In \cite{MBSM12} we have shown that this toroidal organization of the flow is consistent with the tendency of the
velocity field to align with the magnetic field. 
Furthermore, we found that the up-down asymmetry of the geometry causes the generation of a non-zero toroidal angular momentum.
%

\begin{figure}[htb!]
\begin{center}
\includegraphics[width=0.45\linewidth]{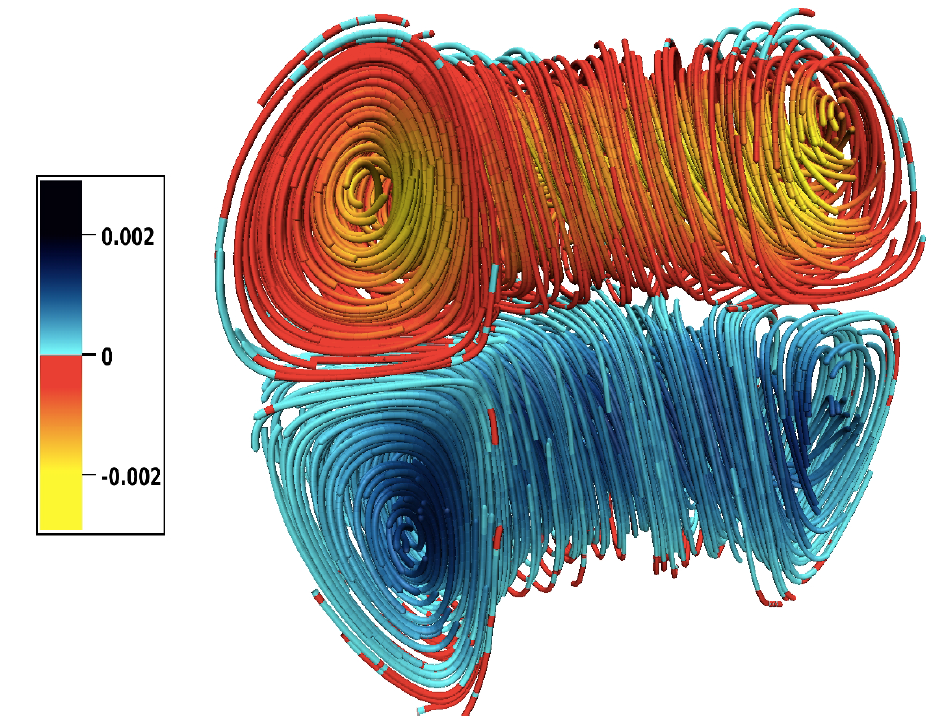}
\includegraphics[width=0.45\linewidth]{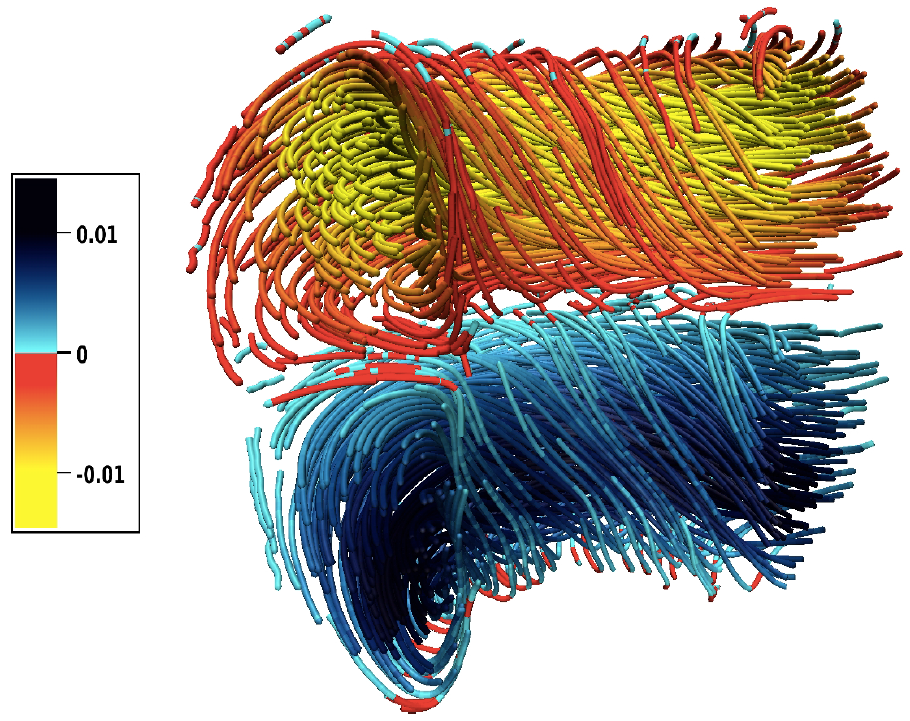}
\end{center}
\caption{\label{fig_torusasym_prl} Confined MHD flows in toroidal domains. Asymmetric poloidal geometry. Streamlines colored with toroidal velocity (uq) for M = 7.5 (left)
and M = 75.2 (right). From \cite{MBSM12}.}
\end{figure}
Figure \ref{fig_torusasym_prl} illustrates the streamlines, colored with the toroidal velocity value for two Lundquist numbers
for the asymmetric poloidal geometry.
For the low Lundquist number case (Fig.~\ref{fig_torusasym_prl}, left) we do indeed observe a pair of counter-rotating vortices in the poloidal plane, while for the larger Lundquist number (Fig.~\ref{fig_torusasym_prl}, right) the flow starts moving in the toroidal direction. A similar behavior is observed for the symmetric poloidal cross-section.
%

%
%
The velocity and the perturbed magnetic toroidal component isocontours at the steady state for a toroidal geometry with asymmetric poloidal cross section are shown in Fig.~\ref{fig_torusasym_jcp} (b and c).
The perturbed toroidal magnetic field is created by the velocities in the poloidal plane.
This component of the magnetic field is important because it generates a toroidal Lorentz
force that induces the toroidal velocities.

\begin{figure}[htbp]
\begin{center}
\includegraphics[width=0.49\linewidth]{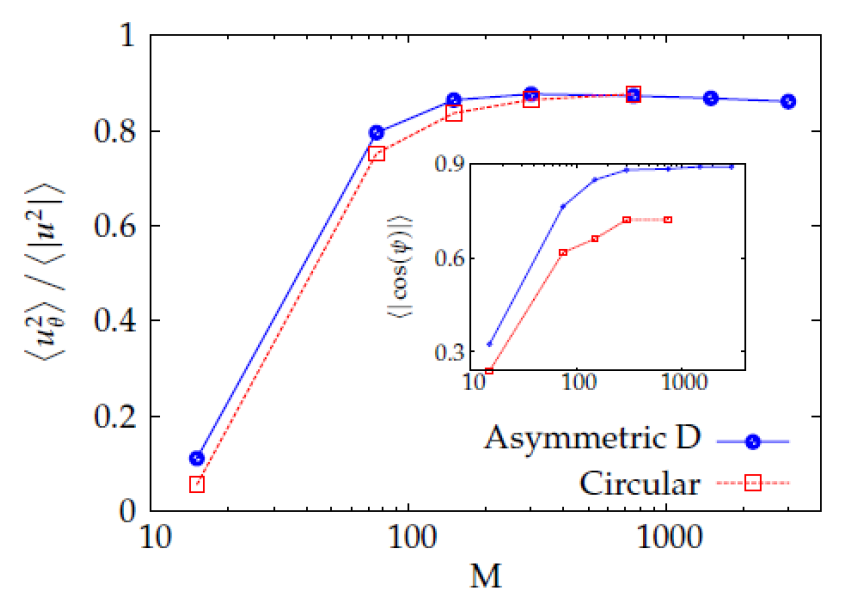}
\includegraphics[width=0.49\linewidth]{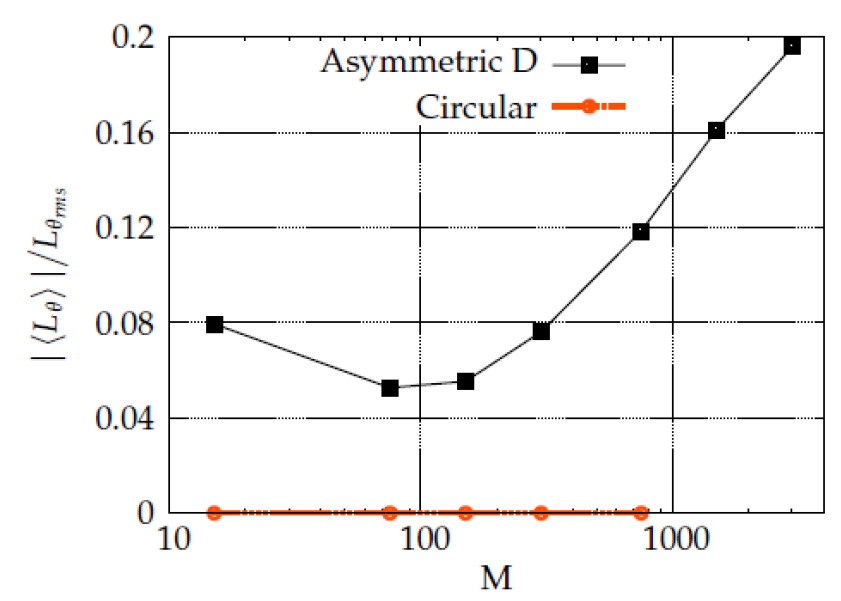}
\end{center}
\caption{\label{fig_torus3d_velocity_moment}
Confined MHD flows in toroidal domains.
Left: normalized toroidal velocity $\langle u_{\theta}^2 \rangle / \langle {\bf u}^2 \rangle$.
The inset shows the modulus of the cosine of the angle between the velocity and the magnetic field.
Right: Volume averaged toroidal angular momentum $\langle L_\theta \rangle $. 
All quantities are plotted as a function of the viscous Lundquist number $M$ for the symmetric and asymmetric poloidal cross section geometries. From \cite{MBSM12}.}
\end{figure}

Figure~\ref{fig_torus3d_velocity_moment} (left) shows that  the toroidal velocity increases with the viscous Lundquist number in both geometries and saturates at about 86 \% of the total squared speed.
The inset shows the modulus of the cosine of the angle between the velocity and the magnetic field and thus quantifies that the velocity fields tend to align with the magnetic field for increasing $M$.
However, for the volume-averaged toroidal angular momentum defined as $\langle L_\theta \rangle \, = \, \frac{1}{V} \, \int_{V} R \, u_\theta \, dV$, we observe fundamental differences for the two geometries as shown in Fig.~\ref{fig_torus3d_velocity_moment} (right).
For the torus with the symmetric poloidal cross-section $\langle L_\theta \rangle $ identically vanishes for all considered Lundquist numbers. In contrast we observe for the asymmetric case that the toiroidal angular momentum increases with the Lundquist number.
%
%
%

\subsection{Effect of toroidicity in RFP dynamics}

\begin{figure}[htbp]
\begin{center}
\includegraphics[width=\linewidth]{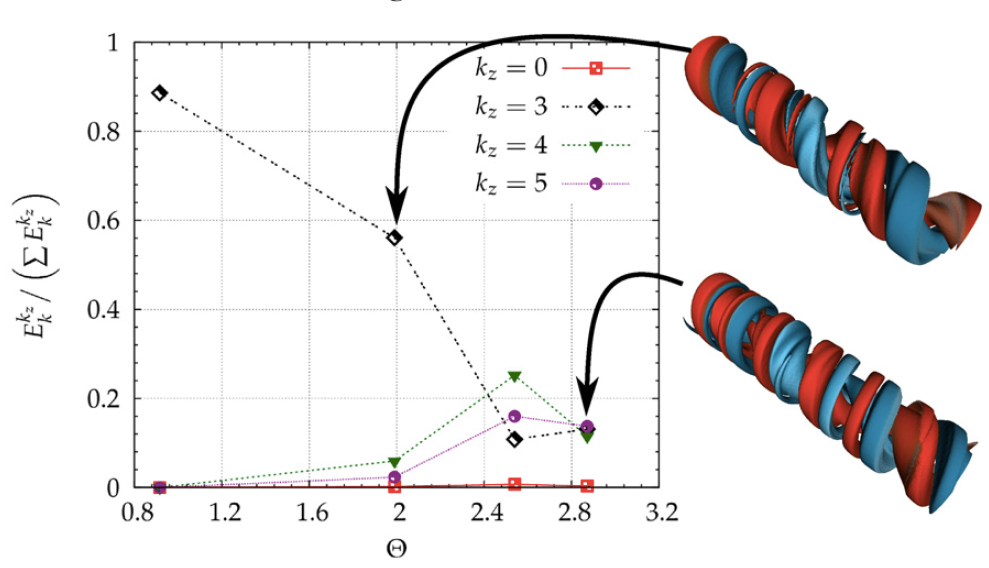}\\
\includegraphics[width=\linewidth]{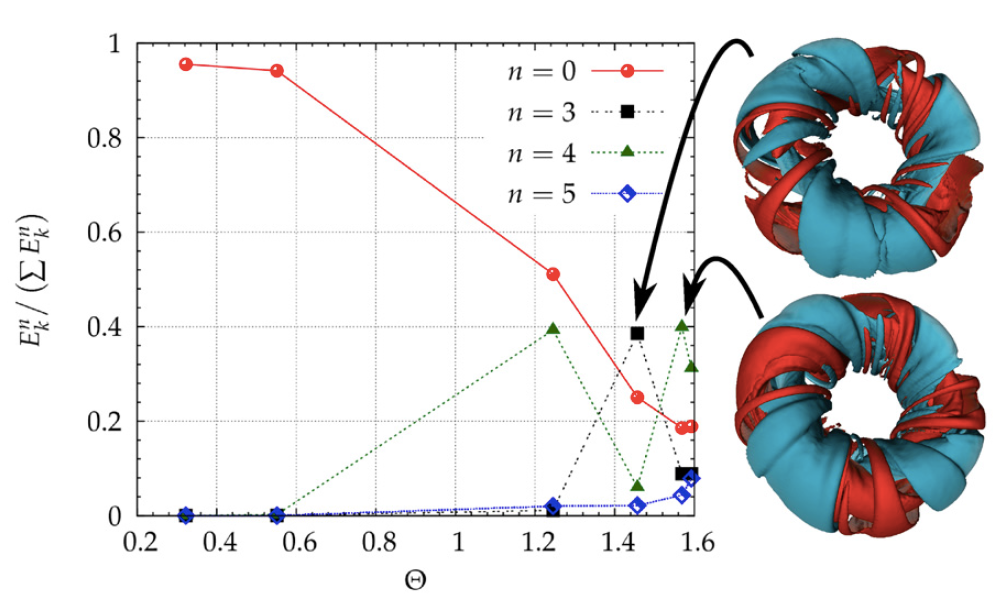}
\end{center}
\caption{RFP dynamics. Ratio of the kinetic energy of the dominant axial (top)/toroidal (bottom) modes over the total kinetic energy for the cylindrical (top) and the torus geometry (bottom) for $M = 888$  as a function of the pinch ratio $\theta$.  
Visualization of the modes: 
Top: axial velocity isosurfaces +0.008 (blue) and −0.008 (orange).
Bottom: toroidal velocity isosurfaces +0.007 (blue) and −0.007 (orange).
From \cite{MBSM14}.}
\label{fig_rfp}
\end{figure}

Finally, we consider the reversed field pinch (RFP) dynamics and study the influence of the curvature of the imposed magnetic field. 
In RFP experiments the plasma evolved to quasi-stationary equilibria characterized by Beltrami minimum energy states
for which the magnetic field corresponds to eigenfunctions of the curl operator \cite{Nord94}.
In  \cite{MBSM14} we compared the RFP flow of a magnetofluid in a torus with aspect ratio $1.83$ with
the flow in a periodic cylinder and studied the persistence of these dominant helical modes for varying pinch ratios. 
%
\begin{figure}[ht!]
\begin{center}
\includegraphics[height=5.5cm, width=0.70\linewidth]{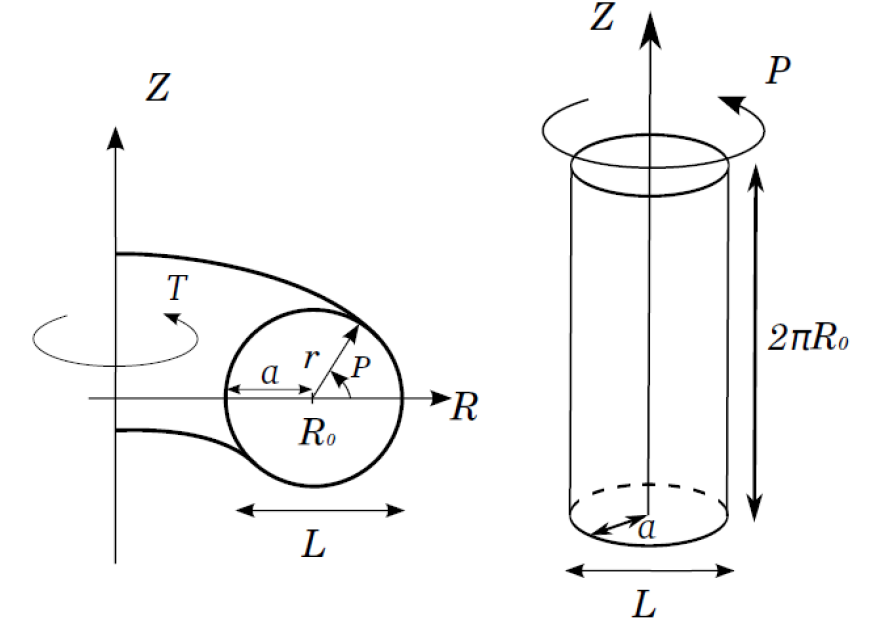}
\end{center}
\caption{Flow configuration of the RFP dynamics. Left: toroidal geometry. Right: periodic cylinder. 
From \cite{MBSM14}.}
\label{fig_rfp_sketch}
\end{figure}
%
The flow configuration of both geometries is illustrated in Fig.~\ref{fig_rfp_sketch}.
The pinch ratio is defined as the wall-averaged poloidal magnetic field
divided by the volume-averaged toroidal magnetic field, $\Theta \, = \, \overline{B_P} / \langle B_T \rangle$ .
The ratios of kinetic energy of the dominant axial and toroidal mode of the total kinetic energy versus the pinch ratio are  shown in Fig.~\ref{fig_rfp} (top) and (bottom) for the cylindrical and the toroidal geometry, respectively.
We find that an axisymmetric toroidal mode is always
present in the toroidal, but absent in the cylindrical configuration. 
In particular, in contrast to the cylinder, the toroidal case presents a double poloidal recirculation cell with a shear
localized at the plasma edge. 
Quasi-single-helicity states are found to be more persistent in
toroidal than in periodic cylinder geometry.
The dominant helical modes at pinch ratios $\Theta \ge 2$ are illustrated in Fig.~\ref{fig_rfp} by showing axial and toroidal velocity isosurfaces to get further insight into the flow topology.
For further details we refer to \cite{MBSM14}.
%

\section{Conclusion and perspectives}

We reviewed immersed boundary methods with a special focus on the volume penalization method
for imposing Dirichlet (corresponding to no-penetration and no-slip conditions) or Neumann boundary conditions (corresponding to no flux conditions) in complex geometries.
The mathematical concepts for choosing the parameters involved, i.e., the penalization parameter and the grid size have been illustrated considering simple one dimensional problems.
Applications to hydrodynamic and magnetohydrodynamic turbulence in complex geometries using a simple Fourier pseudo-spectral method, which can be parallelized on massively parallel machines using standard libraries like P3DFFT, illustrated the versatile use of this technique for various problems encountered in computational physics.
Toroidal geometries can thus be efficiently handled even including asymmetric poloidal cross-sections and simulations for higher Lundquist numbers become feasible.
An essential feature of the volume penalization method is that it becomes more attractive for computing fluid flows for small viscosity values, i.e., for high Reynolds/Lundquist number flows. The reason is that the effective penalization boundary layer size depends on the product of viscosity and the penalization parameter and thus the method becomes more precize without using prohibitively small penalization parameters, cf.~\cite{NKS14}.

One perspective of current research is the development of higher-order penalization methods which allow faster convergence to be obtained. 
The Cartesian grid introduces in dimension larger equal to two a staircase effect for complex  (non grid aligned) geometries and the approximation of the
mask function thus reduces to first order. 
Techniques based on interpolation to obtain higher
order for complex geometries  have been proposed, e.g., in \cite{SVCA08} in the context of finite volume formulations.
Another challenging topic are more sophisticated boundary conditions for MHD flows overcoming the limitation of perfect conductors in the surrounding solid domain.

\section*{Acknowledgements}

K.S. is grateful to Sadri Benkadda for inviting him to give a review lecture at the ITER International School 2014 "High performance computing in fusion science" on which the manuscript is based on.
K.S. is also indebted to Philippe Angot, Wouter Bos, Thomas Engels, Marie Farge, Benjamin Kadoch, Dmitry Kolomenskiy, Matthieu Leroy, Jorge Morales, Salah Neffaa, Romain Nguyen van yen with whom the volume penalization method has been developed together and published in numerous research papers on which this review is based on. 
KS thankfully acknowledges financial support from the ANR project SiCoMHD (ANR-Blanc 2011-045) and thanks
Matteo Faganello for useful comments on the manuscript.


\end{document}